\documentclass[11pt]{article}

\usepackage[american]{babel}
\usepackage[pdftex, pdfauthor={Michael Backes, Philipp von Styp-Rekowsky, Sebastian Gerling}, pdftitle={A Novel Attack against Android Phones}, bookmarks=true]{hyperref}

\title{\textbf{A Novel Attack against Android
Phones}\\ \vspace{0.2cm} \small}
\author{ Michael Backes$^{*\dagger}$, 
    Sebastian Gerling$^{*}$, 
    Philipp von Styp-Rekowsky$^{*}$ 
}

\date{}

\begin{document}
\maketitle

\vspace{-1cm}
\begin{center}
  \textit{$^{*}$ Saarland University, Germany\\
  $^{\dagger}$ MPI-SWS}
\end{center}

\vspace{1cm}

\begin{abstract}
  \noindent In the first quarter of 2011, Android has become the top-selling
  operating system for smartphones. In this paper, we present a novel,
highly critical attack that allows unprompted installation of arbitrary applications
  from the Android Market. Our attack is based on a single
  malicious application, which, in contrast to previously known
  attacks, does not require the user to grant it any permissions. 
\end{abstract}

\section{Responsible Disclosure}
\noindent
We reported this vulnerability to Google on June 20, 2011. In order to give them
time to fix the issue, we removed the main content of this paper and generalized
title and abstract. This document merely serves as a timestamp of discovery.

\section{Hash}
\noindent
The SHA-512 hash of the full report as sent to Google is given below:\\
\\
\texttt{
051426b1794e363544893b7123ba3d15c5d878a9ff736162b85479d063a\\
86940fe2d6280774fd98989cce8b1628d5d9428d0691ee4ffcc2c07da82\\
31ca79af5d}

\end{document}